\documentclass[a4paper,alpha-refs]{aejstyles}

\journal{aej}

\usepackage{graphicx}
\usepackage{siunitx}

\title{The Leiden/ESA Astrophysics Program for Summer Students (LEAPS)}

\author[1,\authfn{1}]{Stefano Bellotti}
\author[1]{Andrew D. Sellek}
\author[1]{Piyush Sharda}
\author[1]{Katarzyna M. Dutkowska}
\author[2]{Ashley Chrimes}
\author[1]{Huub Röttgering}

\affil[1]{Leiden Observatory, Leiden University, Leiden, 2300 RA, PO Box 9513, Netherlands}
\affil[2]{European Space Agency (ESA), European Space Research and Technology Centre (ESTEC), Keplerlaan 1, 2201 AZ Noordwijk, the Netherlands}

\authnote{\authfn{1}bellotti@strw.leidenuniv.nl}

\papercat{Resources \& Activities}

\runningauthor{Bellotti et al.}

\jvolume{00}
\jnumber{0}
\jyear{2025}

\begin{document}

\begin{frontmatter}
\maketitle
\begin{abstract}
International student mobility plays a critical role in shaping future research careers, particularly in highly globalized fields such as astrophysics. The Leiden/ESA Astrophysics Program for Summer Students (LEAPS) offers a 10-week, fully funded research program at Leiden Observatory and the European Space Agency’s ESTEC centre for undergraduate and master's students. Designed to foster early research involvement, LEAPS supports students from diverse academic and cultural backgrounds. Since its inception in 2013, LEAPS has hosted 194 students from over 40 countries. Data collected for 165 participants reveal that over 50\% have progressed to Ph.D. studies, with some members of earlier cohorts already securing competitive international fellowships in astronomy. LEAPS participants have collectively contributed to at least 25 peer-reviewed publications and 13 international conference presentations. LEAPS has contributed successfully in preparing undergraduates for research careers in astrophysics through hands-on experience, mentorship, and scientific exposure. By addressing barriers related to financial means and promoting diversity, the program not only enhances individual career trajectories but also contributes to the broader goal of inclusive academic mobility. Continued efforts are needed to further increase global representation and assess long-term impacts on participants’ scientific careers.
\end{abstract}


\end{frontmatter}

\section{Introduction}\label{sec:intro}

Research shows that a primary motivation for students to undertake education abroad is to improve their career prospects \citep[e.g.][]{Rodrigues2012,Petzold2017}, especially considering that the skills accrued during such programs (for example network, intercultural competences, and personal development) are broadly valued by policy makers and practitioners \citep[e.g.][]{Zimmermann2013,Netz2021}. From an employability point of view, international student mobility has a positive effect \citep[e.g.][]{Teichler2007,Kratz2016} given that it may be symbolic of attractive features such as commitment, willingness to work hard, and adaptability \citep{Jacob2019}. \citet{VanMol2017} showed that certain employers regard mobile students with international experience as equipped with decision-making competence, for instance.

\citet{Rodrigues2013} reported a positive correlation between mobility experience and working abroad, that is, the influence that international mobility has in the development of personal preferences of students towards pursuing international careers \citep[see also][for instance]{Pinto2020}. A recent analysis by \citet{Knutsen2024} shows that students with a university degree conducted fully or partly abroad are 19-22\% and 3-5\% more likely to work abroad than their non-mobile peers. Additionally, international mobility is effective at increasing campus diversity, as well as reinforcing the financial position of higher education institutions \citep{Brooks2011}. Institutions that employ international graduates also benefit from the global competencies, multilingual capabilities, and cross-cultural perspectives they bring, which can strengthen international collaborations and enrich institutional culture. The presence of internationally mobile researchers is thus a key driver of scientific progress and institutional competitiveness on a global scale. This is particularly relevant for fields such as astronomy, which heavily relies on international cooperation and collaboration due to its `universal' nature. 

Here, we present the Leiden/ESA Astrophysics Program for Summer Students (LEAPS), which is the framework for summer research opportunities at Leiden Observatory in collaboration with the European Space Research and Technology Centre of the European Space Agency (ESA/ESTEC). Our aim is to describe the program, highlight its potential impact on the career of past participants, discuss limitations and prospective strategies, and overall contextualise it in terms of international student mobility. The aim of LEAPS is to allow students with diverse talents, cultural and scientific backgrounds, and personal circumstances to undertake astrophysics research in Leiden. In addition, the aim is for the students to build their scientific profile towards a research career. Astrophysics is a highly international discipline and, for many students, LEAPS can represent a first opportunity to experience research beyond their undergraduate institute and on an international level. In this paper, we describe the goal and structure of LEAPS in Sect.~\ref{sec:goal}, and its scientific impact and the positive influence on the participants' career in Sect.~\ref{sec:impact}, and finally we present our conclusions in Sect.~\ref{sec:discussion}.

\section{Goal and Selection Criteria of LEAPS}\label{sec:goal}

LEAPS is a fully-funded, 10-week-long summer research program and it is open to students of all nationalities from outside of Leiden University, therefore promoting inward short-term mobility.\footnote{The LEAPS website is available \href{https://leaps.strw.leidenuniv.nl/}{here}.} The program aims to help participants prepare for potential further research at a higher level in three main ways. Firstly, they build their research experience profile and develop useful skills, including problem solving, coding, communication, and networking. Secondly, the program gives participants an opportunity to experience a full-time research project, potentially in a new research area of astronomy, often unavailable as part of the curricula at their respective institutions. The purpose is to assist the students to make decisions such as whether they wish to pursue a research degree, and if so where and in which field. Finally, conducting research at an international establishments such as Leiden Observatory and ESA/ESTEC is a valuable part of a portfolio for applications for further research (namely a postgraduate research degree).

LEAPS was started in 2013 with an aim to provide world class research opportunities to students from across the world and promote Leiden as a place for graduate studies. It is now one of the longest running, fully-funded and open summer research programs in astronomy in the world (see Table~\ref{tab:tab1}). It has been conducted on a yearly basis with an average participation of 18 students, including an online version during the COVID-19 pandemic in 2020. The organising committee of the program is composed exclusively of astronomy postdoctoral researchers at Leiden Observatory and research fellows at ESTEC, with the participation of one administrative member of Leiden Observatory for logistics organisation. 

Each year different research projects are advertised and their scope spans different areas of astronomy and astrophysics including theory, observations, and instrumentation. This allows a wide variety of research projects to be conducted during LEAPS. Before the student selection process starts, the LEAPS committee collects the title and abstract of the projects from the supervisory teams. These are then advertised on the LEAPS website for the student to choose.

The student selection process is performed by the supervisors of each project and consists of i) a review of the student application and ii) an online interview in English. For the first step, applicants are required to send their curriculum vitae, their academic transcripts, and a motivation letter in which they express preferences for up to two of the available research projects. They are also asked to arrange for a reference letter to be sent. Upon reviewing the submitted material, the supervisors prepare a shortlist of candidates for the subsequent interviews. For each project, the candidate with the best scores is selected for participating in the program. 

During both steps of the selection process, the organising committee provides guidelines as to, for instance, making sure that the eligibility criteria are met by the candidates, understanding confidentiality of the material, conducting interviews effectively (with the option to involve external experts), and reducing intrinsic bias in the selections. As of 2023, we introduced a country grading system during the selection process. This system accounts for both the nationality of applicants and their university affiliation at the time of application, considering potential biases and disparities related to access to education, research opportunities, and financial constraints. By evaluating countries based on key developmental and educational factors, such as Human Development Index (HDI\footnote{see the Human Development Reports \href{https://hdr.undp.org/}{website}}), STEM representation\footnote{see the Organisation for Economic Co-operation and Development (OECD) Science, Technology, and Industry Scoreboard \href{https://www.oecd.org/}{website}}, gender and diversity representation\footnote{see the Global Education report at the UNESCO \href{https://www.unesco.org/en}{website}}, and scientific publications\footnote{Quantified as the number of peer-reviewed papers produced by a researcher of a certain country to indicate the level of scientific activity, see also the Scopus \href{https://www.scopus.com/}{website}.}, we aim to provide a fairer way to contextualize candidates' achievements and increase diversity in the program.

Besides their day-to-day project work, as part of the summer program, the students attend presentations and lectures dedicated to, for instance, academic writing and efficient communication in presentations. These are typically given by professional astronomers at Leiden Observatory. Additionally, the LEAPS committee arranges at least two guided visits during the summer program: one visit to ESTEC and one visit to ASTRON (the Netherlands Institute for Radio Astronomy). At the end of the summer program, the LEAPS committee organises a one-day event where every student presents their results to the members of Leiden Observatory. Many students subsequently adapt their presentation for a talk at their home institution or an international conference (see Sect.~3). Overall, these experiences aim to provide and improve basic skills for a career in astronomical research.

\section{Statistics and Impact of LEAPS}\label{sec:impact}

The aim of this paper is to quantify and discuss the impact of the program, hence we have collected information regarding the career status of the participants as of 2023, their country of affiliation at the moment of application, and any scientific output. With the latter, we encompass both the publication of peer-reviewed scientific articles as well as participation in international astronomy conferences where they have delivered a talk or presented a poster. 

We retrieved the data on the country of affiliation from the archives of the program. For the career path and the scientific output, we contacted the students that participated in the program. In case this was not possible, we performed an online search (via LinkedIn, personal websites, University websites, etc). The total number of students who participated in LEAPS between 2013 and 2023 is {194}; and we collected information for 165 of them, resulting in an 85\% completeness. The incompleteness stems from a lack of information in our records or unavailable information from online searches. 

\begin{figure*}[!t]
\centering
\includegraphics[width=\textwidth]{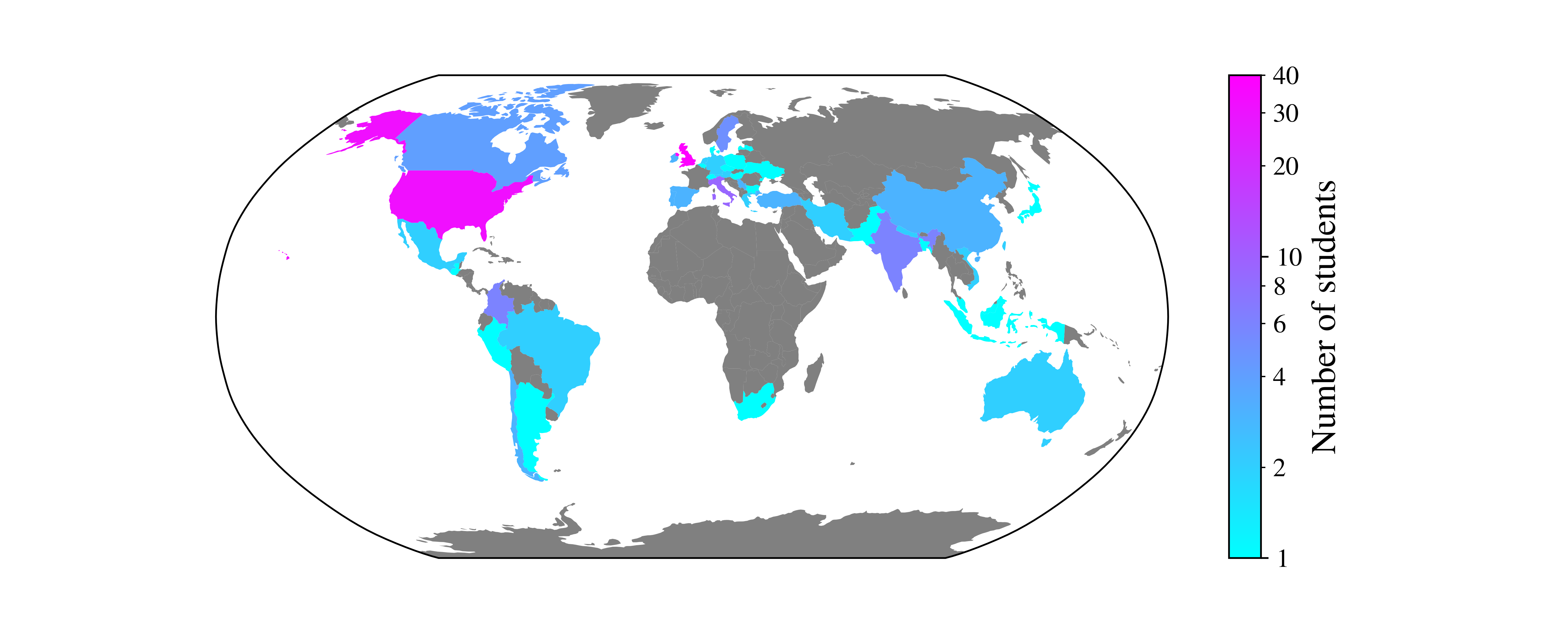}
\caption{Country of affiliation at the moment of application for students that participated in LEAPS between 2013 and 2023. The countries are colour-coded based on the number of students affiliated with that specific country.}\label{fig:countries}
\end{figure*}

LEAPS has attracted applicants from more than 89 countries, with consistently increasing number of applicants per available project over the years. For instance, the oversubscription ratio per project has reached higher than 80 for several projects in the last few years, which highlights the popularity and competitiveness of the program. The number of students with available information on their affiliation is 163. As of 2023, successful candidates have come from institutions in at least 40 countries, as shown in Fig.~\ref{fig:countries}. Around 24\% of the students had an affiliation in the United Kingdom, 20\% from the United States of America, and 6\% from Italy. Each of the remaining countries account for at most 4\% individually. When compared in terms of continent, 53\% of the students had a European affiliation, 23\% a North American one, 9\% a South American one, 13\% an Asian one, and 2\% between Africa and Oceania.

\begin{figure}[!t]
\centering
\includegraphics[width=\columnwidth]{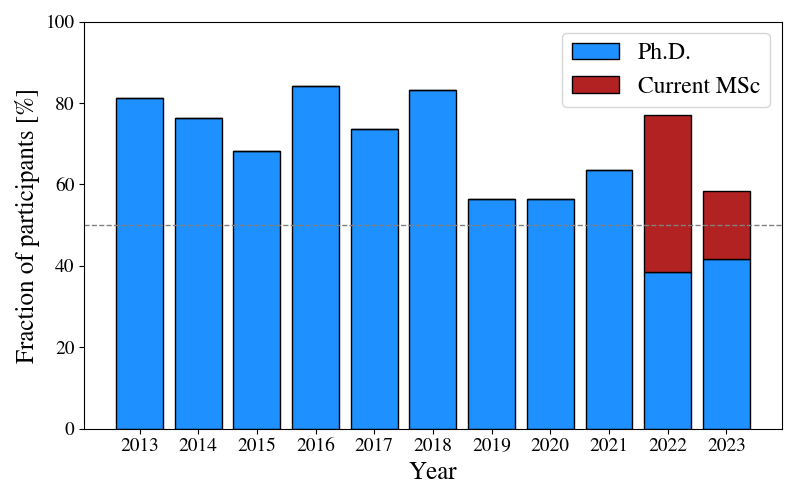}
\caption{Number of students each year who have subsequently undertaken a Ph.D. program normalised by the total number of participants for that year. The red bars for 2022 and 2023 indicate the students who, after LEAPS, were enrolled in a masters program at the time of our survey.}\label{fig:phds}
\end{figure}

Participating in LEAPS may have motivated the students to progress with research and develop prominent international careers. We therefore define our first metric as the fraction of students from each cohort who have since undertaken or are now undertaking a Ph.D. (Fig.~\ref{fig:phds}). In this light, our results are more sensitive toward the short-term impact of LEAPS rather than a long-term one; considering the timescale for which the program has been running, only a small subsample of students would have had time to reach more senior positions.

In total, 131 former LEAPS students have subsequently undertaken a Ph.D., 67\% of the total number.
From Fig.~\ref{fig:phds}, we note that consistently, more than 50\% of LEAPS students secured a Ph.D. position and that there are no long-term trends in the Ph.D. fraction. The years 2022-2023 see a low number of students conducting a Ph.D. simply because most participants were in a master's program at the time of our survey.
Overall, this suggests that indicating that our conclusions on the Ph.D. propensity are not biased by the completeness of our sample.

Moreover, we further consider the mobility of those students in the Ph.D. studies by determining the fraction who moved country in that process. The number of LEAPS participants that have undertaken or are undertaking a Ph.D., and for which we have information about their affiliation both before and after LEAPS, is 122. Of them, we recorded 74 students (61\%) choosing a different country for the Ph.D. than that of the affiliation before LEAPS. Instead, 48 students (39\%) chose the same country.  Overall, LEAPS may have provided additional motivation to pursue an international career and thus increase student international mobility.

We include here a few testimonials from LEAPS students over the years, in order to provide more qualitative data on the influence of LEAPS on their skills, network, and motivation. 
\begin{itemize}
    \item S. Yen (LEAPS 2013) -- {\it `I'm glad to hear LEAPS continues after all these years. It was such a valuable and insightful experience. I absolutely loved it!'}
    \item T. Boztepe (LEAPS 2018) -- {\it `I learned various things, including programming skills, astronomy, academic culture, etc. from [supervisor]. During my time in LEAPS, [supervisor] was patient and helpful and helped me to improve myself.'}
    \item L. Pham (LEAPS 2021) -- {\it `The experiences I gained from LEAPS are significantly helpful for me to overcome challenges in research.'}
    \item G. Pignataro (LEAPS 2021) -- {\it `LEAPS actually helped me get into the PhD program at the University of [city], the same year.'}
    \item D. Nadella (LEAPS 2022) -- {\it `It was a valuable experience and we were fortunate enough to write a paper about our research that summer.'}
\end{itemize}

In turn, LEAPS may have contributed to establishing Leiden as a sought-after destination for graduate studies in astronomy, with applications to the Leiden Observatory Ph.D. program increasing sharply since the program's inception. From the available information about the Ph.D. affiliation after LEAPS, we found that Leiden Observatory is the Ph.D. institute of twelve students, with an average of one student per year. In a couple of cases, these Ph.D. students have given back to the program, such as by appearing on a careers panel or even acting as a project mentor to their own student. Moreover, many LEAPS alumni have successfully obtained prestigious fellowships around the world that are noteworthy in astronomy: one NASA Sagan fellowship, one Hubble fellowship, one Einstein fellowship, one 51 Pegasi b fellowship, one Humboldt fellowship, two ESO fellowships, and one De Sitter fellowship.

For the scientific output of LEAPS, we found that at least 25 research papers have been published in peer-reviewed astronomy journals\footnote{The list of output from LEAPS is available here \href{https://ui.adsabs.harvard.edu/search/q=docs(library\%2F4Y2oVuDvRxCfCXtmt6ZQqw)\&sort=date\%20desc\%2C\%20bibcode\%20desc\&p\_=0}{LEAPS ADS Library}}. Similarly, the astrophysics data system (ADS) records a conference participation in the form of a talk or a poster for at least 13 additional projects (we note that only some conferences record their contributions in that database). Overall, these numbers highlight the successful research productivity from LEAPS, the improvement of the student's curriculum to pursue further research, and the establishment of longer-term working relationships beyond the ten weeks of the program itself. Not only does LEAPS benefit the students, but also the early-career researchers who devise and supervise the projects and run the program. For instance, early-career researchers obtain hands-on experience on shortlisting candidates and managing selection interviews, also helping them to reflect on their own applications and those of any other students they advise. The mentoring and coordinating experiences are valuable for their career development and progression.

\begin{table*}[!t]
\centering
\caption{Comparison of LEAPS with other summer research programs in astronomy in the world. Fully funded indicates the inclusion of accommodation, travel, and a stipend.}\label{tab:tab1}
\begin{tabular}{|l|l|c|c|c|c|c|}
\hline
{\bf No.} & {\bf Program} & {\bf Organizer} & {\bf Eligible Countries} & {\bf Duration} & {\bf Fully Funded} & {\bf Inception}\\
\hline
1 & RISE Germany & German Government & 4 Countries & 12 weeks & Yes & 2005 \\
2 & RISE Worldwide & German Government & Germany & 12 weeks & Yes & 2010 \\
\hline
3 & Globalink & MITACS Inc. Canada & 17 Countries & 12 weeks & Yes & 1999 \\
\hline
4 & SURP & University of Toronto & Canada & 4 months & Yes & N/A \\
\hline
5 & SASP & STScI USA & All & 9 weeks & Yes & N/A \\
\hline
6 & ASTRP & Oxford University & UK & 8 weeks & Yes & N/A \\
\hline
7 & REU & NSF USA & USA & 10 weeks & No & 1987 \\
\hline
8 & FRT & Australian National University & India, Indonesia & 10-12 weeks & Yes & 2019 \\
\hline
9 & SRPs & Australian Universities & Australia, New Zealand & 6-12 weeks & Yes & Variable \\
\hline
10 & ESO-SRP & European Southern Observatories & Prefer ESO & 6 weeks & Yes & 2019 \\
& & Member States & Member States & & & \\
\hline
11 & RASS & Raman Research Institute & India & 8 weeks & Yes & N/A \\
\hline
12 & Astro Camp & Universidade do Porto & 42 Countries & 2 weeks & No & 2012 \\
\hline
13 & SSP & ASIAA Taiwan & All & 8 weeks & Yes & 1998 \\
\hline
14 & VOSS & Vatican Observatory & All & 4 weeks & Yes & 1986 \\
\hline
15 & ASPIRE & University of Amsterdam & All & 6 weeks & Yes & 2015 \\
\hline
16 & SRP & ASTRON Netherlands & All & 10-12 weeks & Yes & 2015 \\
\hline
17 & DAWN-IRES & NSF USA & USA & 11 weeks & Yes & 2019 \\
\hline
18 & RECA & RECA Association & Colombia & 10 weeks & No & 2021 \\
\hline
19 & LEAPS & Leiden University \& ESA/ESTEC & All / prefer ESA Member & 10 weeks & Yes & 2013 \\
 &  &  & States for ESTEC &  & &  \\
\hline
\end{tabular} \\
\end{table*}

\section{Summary and Discussion}\label{sec:discussion}

In this paper, we introduced the LEAPS program for summer student conducted jointly by Leiden Observatory and ESA/ESTEC. We described the impact of the program between 2013 and 2023 in terms of current career status of the participants, their country of affiliation at the moment of application, and their scientific output. The information collected from the database of LEAPS shows an overall positive contribution, with at least 50\% of the participants pursuing doctoral studies from almost all years of LEAPS, 25 peer-reviewed published papers, and at least five world-renowned research fellowships. These aspects indicate that LEAPS may successfully contribute to provide basic skills and competences for a career in astronomical research. With participants from at least 40 countries, LEAPS is providing the international study experience that may, given the findings of \citet{Knutsen2024}, help these students secure a career abroad should they wish.

Generally, there are certain challenges to international student mobility to be taken into account \citep{Brooks2024}: student well-being, financial support, and diversity. The student well-being is an aspect that LEAPS takes seriously because, in the context of academic internationalisation, it is known that carrying out a research project in a foreign country may be stressful. Ultimately, this could lead to a negative influence on the satisfaction with the experience abroad \citep{Ramirez2024}. During LEAPS, each project must have two supervisors to i) safeguard the welfare of the student, ii) make sure that at least one supervisor is present for the duration of the program for regular updates on the student’s progress, and iii) allow the student to benefit from wider scientific expertise. In addition, the students have access to welfare channels such as Observatory Allies and Confidentiality Counsellors, and the LEAPS organising committee performs a welfare check around halfway through the program, to address whether the students are well-settled in Leiden, and are satisfied with their living situation, healthcare, sufficiency of the stipend, relationship with their supervisors, and provision of project support. The camaraderie of the LEAPS cohort, nurtured by housing the students together and having them share an office, is the final pillar of their social support. Every LEAPS cohort has been seen to independently organize travel and trips on the days off to explore life in Europe. LEAPS alumni also report remaining in contact over the years, and continue to support each other in various ways.

Another challenge to face is the affordability of international students mobility \citep{King2010,Souto-Otero2013,Brooks2024}, which most likely impacts students from less privileged social backgrounds \citep[see e.g.][]{Slowey2020}. To help navigate the costs of living abroad, LEAPS provides a complete package. In practice, the costs of travel to/from the Netherlands, accommodation for the duration of the program, international student insurance, visa applications, and living expenses (in the form of a stipend) are covered by the program. This aspect of the program ensures an equitable ability for students to partake in research opportunities regardless of the financial status. The funding for LEAPS has been periodically increased to take inflation into account, and to keep it on par with other summer programs in the Netherlands. Table \ref{tab:tab1} shows that, to the best of our knowledge, LEAPS is now the longest running, fully-funded astronomy summer program that runs for an entire summer (10 weeks) and is open to citizens of all countries. In particular, the emphasis of LEAPS on keeping it open to every eligible student in the world has enabled a uniquely diverse cohort of students from across the globe to engage in frontline research, fostering an international and inclusive training environment.
A more quantitative comparison of the impacts of LEAPS and the other astronomy summer programs is not possible as, to our knowledge, no comparable program has aggregated and published similar statistics to those presented herein.

An area of improvement of the LEAPS program lies in the diversity of the students. Although LEAPS strives to ensure fairness by mitigating any financial disadvantage of the participants, and the LEAPS selection process involves multiple supervisors at every step, in order to increase fairness,  Fig.~\ref{fig:countries} illustrates that the majority of students were affiliated with European and Northern American countries, which tend to have larger, more internationally recognized academic systems, and more scientific outputs. Notably, the countries with the greatest number of previous participants also provide the most applicants, suggesting a level of self-reinforcement: in these countries the program becomes well-known and previous participants or their advisors mentor further applicants to success. Therefore, different strategies are needed in order to increase diversity towards students that are traditionally less likely to engage in international mobility, for example through increased promotion of the program in under-represented areas. As mentioned in Sect.~\ref{sec:goal}, we recently developed a country grading system aimed at reducing disparities among applicants, but future monitoring is required to evaluate the effect of its implementation on the diversity of LEAPS participants. 

An interesting case study that emerged from our data is Colombia. The number of LEAPS participants affiliated with this country is outsized with respect to its i) population and ii) scientific strength, as referred to based on parameters like HDI, STEM representation, and scientific publications (see Sect.~\ref{sec:goal}). This could plausibly be due to an effective advertisement of LEAPS by previous students from Colombia or during the AstroTwinCoLO programme\footnote{Information about the programme can be found on the \href{https://sites.google.com/site/astrotwincolo/home?authuser=0}{website}}. The latter is a twinning programme between the University of Leiden and the University of Antioquia aimed at boosting the development and improving the quality of astronomy teaching and research at the University of Antioquia and the Andean region.

For further context, we can compare to the statistics on the astronomy-related doctorates awarded in each country per million residents in \citet[][Table~3]{Tenn2023}. We observe that only six individuals ever have obtained a Ph.D. in astronomy in Colombia (0.1 astronomy-related Ph.D.s per million). From our database, we have six students from Colombia who participated in LEAPS, three of whom pursued doctoral studies in the Netherlands after LEAPS (for the remaining three, we have no information that they pursued a Ph.D.). Given this significant number, LEAPS, and the international mobility the students gain from it, may have significantly increased the opportunity for those students to pursue a Ph.D. compared to the opportunities available in their home country. For the other countries appearing in Table~3 of \citet{Tenn2023}, and for which we have information of more than three students in our database (namely Canada, Sweden, United Kingdom, and United States of America; 30.2-83.9 astronomy-related Ph.D.s per million), we cannot draw similar conclusions. This is because our numbers are only a small fraction of the number of Ph.D. listed in Table~3 of \citet{Tenn2023}, i.e. students from those countries already have strong access to Astronomy PhD programs. All other countries that appear in both their Table~3 and our data base have less than three LEAPS participants, so any claim would be biased by low-number statistics. However, combined with the fact that the 50\% rate of LEAPS students from Colombia going on to undertake Ph.D.s is statistically indistinguishable from the overall rate of 70\%, and the 59\% rate for students from the United States of America (50.8 astronomy-related Ph.D.s per million), this suggests that LEAPS may be helping to level the playing field for students from less scientifically advantaged countries.

We note that the focus of LEAPS is on international mobility, thus here we have highlighted geographic imbalance. Other equally important dimensions of diversity such as gender are not explicitly targeted by the program, and correspondingly the required data to evaluate these aspects are not always recorded. Nevertheless, we note that in all years with available data, a majority of participants were female. This suggests that current advertising and selection practices are sufficient to maintain participation from those historically marginalized on account of gender.

Overall, the idea behind LEAPS is to train students to perform hands-on astronomical research and develop the basic scientific skillset to conduct further research. Such a concept is flexible and adaptable to different scenarios. For instance, programmes such as LEAPS could be down-scaled and adapted to high-schools. The idea would be to propose simple research projects to students, eventually not for a single student but in groups, with the aim to mentor them and teach them how to conduct scientific research. While the international mobility aspect may not be applied directly for high-school programmes, it may apply at national level, depending on the country. The study of \citet{Salto2014} has shown that similar programmes at high-school level can boost the long-term motivation to engage in research careers.

In conclusion, the astronomy internships available worldwide have become very competitive even at the undergraduate and masters stage, given the steep rise in astronomy research around the world. This demonstrates the success of several efforts aimed at astronomy outreach in high schools and colleges around the world. However, future programs should look at sustainability in terms of how to balance the vast number of talented and interested applicants worldwide with the small number of positions available.

\section{Declarations}

\subsection{List of abbreviations}

ADS = Astrophysics Data System\\
ESA = European Space Agency\\
ESO = European Southern Observatory\\
ESTEC = European Space Research and Technology Centre\\
HDI = Human Development Index\\
NSF = National Science Foundation\\
STScI = Space Telescope Science Institute\\
ASIAA = Academica Sinica Institute of Astronomy and Astrophysics\\
RECA = Red de Estudiantes Colombianos de Astronomía\\
LEAPS = Leiden/ESA Astrophysics Program for Summer students\\
OECD = Organisation for Economic Co-operation and Development\\
STEM = Science, Technology, Engineering, and Mathematics

\subsection{Consent for publication}

Not applicable.

\subsection{Competing Interests}

The authors declare that they have no competing interests.

\subsection{Funding}

S. Bellotti acknowledges funding by the Dutch Research Council (NWO) under the project "Exo-space weather and contemporaneous signatures of star-planet interactions" (with project number OCENW.M.22.215 of the research programme "Open Competition Domain Science- M"). A. D. Sellek acknowledges support from the ERC grant 101019751 MOLDISK. P. Sharda is funded by the Leiden University Oort Fellowship and the International Astronomical Union Gruber Fellowship. K. M. Dutkowska acknowledges support from the European Research Council (ERC) Advanced Grant MOPPEX 833460. A. Chrimes acknowledges support through the European Space Agency (ESA) research fellowship programme. We would like to thank S. Yen, T. Boztepe, L. Pham, G. Pignataro, and D. Nadella for providing testimonials of their LEAPS experience.

\subsection{Author's Contributions}

S. Bellotti, A. Sellek, P. Sharda, K.M. Dutkowska, and A. Chrimes were the members of the LEAPS committee when this paper was conceptualised. S. Bellotti curated and analysed the data. S. Bellotti, A. Sellek, P. Sharda, and K.M. Dutkowska provided resources and references for the paper, and wrote the original draft of the paper. S. Bellotti and P. Sharda created the plots and tables for visualising the results. K.M. Dutkowska designed the country grading system implemented during the LEAPS selection process. All authors participated in the editing and review of the paper.

\section{Acknowledgements}

We would like to thank the anonymous referees for their constructive feedback and helpful suggestions. We would like to thank Adam Muzzin for spearheading the initiation of LEAPS, and Ignas Snellen, Jos de Bruijne, and Gaitee Hussain for supporting LEAPS over the years. We would like to thank Erik Deul for the outstanding IT support and the development and maintenance of the registration platform over the years, allowing LEAPS to run smoothly. We also thank the various staff members of Leiden Observatory who have financially supported LEAPS students from their research grants. We would like to thank Monica Lamers and Nancy Zhou for the administrative work behind the smooth organisation of the summer programme. We would also like to thank all past and present members of the LEAPS committee and the supervisors for the organisation and commitment to LEAPS. We would finally like to thank the former LEAPS participants who provided us with information on their subsequent career paths.



\bibliography{paper-refs}

\end{document}